\documentclass[final,12pt]{elsart}

\usepackage{amsmath,amssymb} 
\usepackage[dvips]{graphicx} 

\def\figurename{Fig.}
\def\tablename{Tab.}

\def\epsilon{\varepsilon}
\def\phi{\varphi}

\newcommand{\be}{\begin{equation}}
\newcommand{\ee}{\end{equation}}
\newcommand{\ba}{\begin{array}}
\newcommand{\ea}{\end{array}}
\newcommand{\bea}{\begin{eqnarray}}
\newcommand{\eea}{\end{eqnarray}}
\newcommand{\pro}{\partial}

\newcommand{\oneg}{\displaystyle\frac{1}{g}}
\newcommand{\valpha}{\vec \alpha}

\newcommand{\hn}{{\hat n}}

\newcommand{\difrac}{\displaystyle\frac}
\newcommand{\nn}{\nonumber}

\newcommand{\astK}{\stackrel{\ast}{K}}
\newcommand{\TS}{\stackrel{\top}{S}}
\newcommand{\TTS}{\stackrel{\top\!\!\top}{S}}
\newcommand{\TR}{\stackrel{\top}{R}}
\newcommand{\TTR}{\stackrel{\top\!\!\top}{R}}
\newcommand{\D}{{\hat D}}
\newcommand{\X}{{\vec X}}
\begin{document}
\begin{frontmatter}
\title{Confinement, Vacuum Structure: from QCD to Quantum Gravity} 
\author[First]{D. G. Pak} 
\address[First]{Center for Theoretical Physics, Seoul National
University, Seoul 151-742, Korea}
\ead{dmipak@phya.snu.ac.kr}
\begin{abstract} 
A minimal Lorentz gauge gravity model with $R^2$-type Lagrangian is proposed.
In the absence of torsion the model admits a topological phase
with unfixed metric.
The model possesses a minimal set of dynamical degrees of freedom
for the torsion. Remarkably, the torsion has the same
number of dynamical of-shell degrees of freedom as the metric tensor.
We trace an analogy between the structure of the quantum chromodynamics
and the structure of possible theory of quantum gravity.
\end{abstract}
\end{frontmatter}
\section{Introduction}
The gauge approach to gravity based on \cite{uti}
gauging the Lorentz and Poincare groups \cite{uti, kibble,cho11}
was proposed as a possible way to construct
a consistent quantum theory
of gravity. The extension of gravity models to the case
of non-Riemannian space-time geometry reveals new possibilities
towards construction of renormalizable quantum gravity with torsion
\cite{ivan1,hehl}. 
Recently, a Lorentz gauge model of gravity with Yang-Mills type Lagrangian
including torsion has been developed further in \cite{pak}.
It has been proposed that the Einstein gravity
can be induced as an effective theory via mechanism similar to
the dual Meissner effect of color confinement in 
quantum chromodynamics (QCD).
In that model the space-time metric is treated as a fixed classical
field while the contortion supposed to be a quantum field.
Such a treatment of the metric is not satisfactory from the
conceptual point of view since one has to assume the existence
of the classical space-time with a metric given a priori. 
In the present paper we propose 
a model which admits the existence
of a pure topological phase with an arbitrary metric from the start.
We conjecture that the torsion can be confined in analogy
to confined gluons in QCD.

\section{Abelian projection in $SU(3)$ QCD}

Let us start from the concept of the Abelian projection in QCD \cite{cho1}. 
The principal role in this construction belongs to the scalar
fields $\hn^a_i, a=1,2,3 i=1,2$ which parameterize the coset $SU(3)/U(1)\times U(1)$.
In the general construction of the Abelian projection 
\cite{ periw,fadd1} the scalar field $\hn$ is given by a set 
of over determined variables which is not convenient 
for description of the effective theory like Faddeev-Niemi-Skyrme 
model \cite{fadd2}.
We give an explicit construction of the Abelian projection
for the group $SU(3)$ with a minimal set of degrees of freedom
for $\hn^i$. 

The Cartan algebra of $SU(3)$ Lie algebra is generated
by two vectors ${\bf n_{3}} = \hn_{3}^a t^{3} \, ,
{\bf n_{8}} = \hn_{8}^a t^{8}$ with
$t_{3,8}$ as generators of $SU(3)$.
Let us parametrize the lowest wight vector $\hn_8^a$ in terms of complex 
triplet field $\Psi$ which parameterized
the coset $  CP^2 \simeq SU(3)/SU(2)\times U(1)$
\bea
&& \hn_8^a = N_1 \bar \Psi \lambda^a \Psi, ~~~~~~~~~~~ \bar \Psi \Psi=1 , \label{par1}
\eea
where the normalization factor $N_1=-3/2$ provides  
the conditions 
\bea
&& \hn_8^2=1, ~~~~~~~~d^{abc}\hn_8^b\hn_8^c=-\difrac{1}{\sqrt 3} \hn_8.
\eea
To construct the second Cartan vector $\hn_3$, which is orthogonal
to $\hn_8$, it is convenient to define projectional
operators
\bea
&& P_{\parallel}^{ab}=\hn_8^a \hn_8^b ,~~~~~~~~~~ P_\bot^{ab} = \delta^{ab} - \hn_8^a \hn_8^b
\eea
and introduce another independent complex triplet field $\Phi$ ($\bar \Phi \Phi =1$)
orthogonal to $\Psi$.
With this the vector $\hn_3$ can be parameterized as follows
\bea
 \hn_3^a &=& P_\bot^{ab} \bar \Phi \lambda^b \Phi
=\bar \Phi \lambda^a \Phi +\difrac{1}{2} \bar \Psi \lambda^a \Psi .\label{par2}
\eea
The parametrization defined by (\ref{par1}, \ref{par2})  
is invariant under dual local $\tilde U(1)
\times \tilde U'(1)$ group transformation.
The Abelian projection of $SU(3)$ gauge connection
is similar to the decomposition of $SU(2)$ gauge
potential \cite{cho1}
\bea
&& \vec{A}_\mu =\hat A_\mu +\vec C_\mu 
+ \vec X_\mu, ~~~~~~~~
\hat A_\mu = A_{\mu i} \hn_i + \vec C_\mu \nn \\
&& \vec C_\mu^a =-f^{abc} \hn^{bi} \pro_\mu \hn^{ci} \equiv
 -\hn_i \times \pro_\mu \hn_i,  \label{cdec}
\eea
where $\hat A_\mu$ is a restricted potential, 
$\vec C_\mu$ is a magnetic potential,
and $\vec X_\mu$
represents the off-diagonal (valence ) gluon.
One can verify that 
the vectors $\hn_i$ are covariantly constant
\bea
&& \hat D_\mu \hn_i \equiv (\pro_\mu + \hat A_\mu) \hn_i=0.
\eea
The decomposition (\ref{cdec}) allows two types of gauge 
transformation: (I) the background gauge transformation described by
\bea
&\delta \hat A_\mu = \oneg \D_\mu \valpha,
~~~~~\delta \X_\mu = - \valpha \times \X_\mu, 
\label{bgt}
\eea
and, (II) the quantum gauge transformation described by
\bea
&\delta \hat A_\mu = 0,
~~~~~&\delta \X_\mu =\oneg \D_\mu \valpha.
\label{qgt}
\eea
The background gauge transformation shows that $\hat A_\mu$ 
by itself satisfies the full $SU(3)$ 
gauge degrees of freedom, even though it describes the Abelian
part of the potential. Furthermore
$\vec X_\mu$ transforms covariantly like a vector.
\section{Parallels between QCD and Quantum Gravity}

The basic geometric objects in approaches to formulation of gravity as a gauge
theory of the Poincare group \cite{uti,kibble,cho11} are
the vielbein $e_a^m$ and the general Lorentz affine connection $A_m^{~cd}$.
The covariant derivative with respect to Lorentz gauge
transformation is defined in a standard manner
\bea
D_a=e_a^m (\pro_m + {\bf A}_m) ,
\eea
where ${\bf A}_m\equiv A_{m cd} \Omega^{cd}$ is a general
affine connection taking values in the Lorentz Lie algebra.
The affine connection $A_{m cd}$ can be rewritten
as a sum of Levi-Civita spin connection $\varphi_{m c}^{~~d}(e) $
and contortion $K_{m c}^{~~d}$
\bea
A_{m c}^{~~d} &=& \varphi_{m c}^{~~d}(e) + K_{m c}^{~~d}. \label{split}
\eea

In analogy with QCD we can define two types of Lorentz
gauge transformations consistent with the
original Lorentz gauge transformation: \newline
(I) the classical, or background, gauge transformation
\bea
&&\delta e_a^m = \Lambda_a^{~b} e_b^m, ~~~~~~~ \delta {\boldsymbol \varphi}_m(e) = -\pro_m {\bf \Lambda}-
                 [{\boldsymbol \varphi}_m,{\bf \Lambda}], \nn\\
&& \delta {\bf K}_m = -[{\bf K}_m,{\bf \Lambda}],\label{eqI}
\eea
(II) the quantum gauge transformation
\bea
&&\delta e_a^m =
\delta {\boldsymbol \varphi}_m(e)=0, ~~~~~ \delta {\bf K}_m= - \hat D_m {\bf \Lambda}-[{\bf K}_m,{\bf \Lambda}],\label{eqII}
\eea
where ${\boldsymbol \varphi}_m \equiv \varphi_{m cd} \Omega^{cd}$, and
the restricted covariant derivative $\hat D_m$ is defined
by means of the Levi-Civita connection only.
Under the decomposition (\ref{split})
the Riemann-Cartan curvature is splitted into two parts respectively
\bea
R_{abcd}=\hat R_{abcd}+\tilde R_{abcd}.
\eea

From the comparison of the Abelian decomposition in QCD 
with the decomposition of the Lorentz spin connection
one can find the
analogy between the restricted potential $\hat A_\mu$ and valence
gluon $\vec X_\mu$ in QCD on the one hand,
and the Levi-Civita connection $\omega_{\mu cd}$
and contortion $K_{\mu cd}$ in Lorentz gauge gravity
on the other.
Obviously, in QCD we can not treat the off-diagonal
component $\vec X_\mu$ as a true vector. The reason is that
if we introduce, for instance, a mass term for the off-diagonal gluon
into the Lagrangian then the renormalizability will be lost.
For the same reason we can not treat the contortion $K_{\mu cd}$
as a true tensor in gravity models in attempts
to formulate a quantum renormalizable theory in the case if we wish
to keep two types of Lorentz  gauge symmetries.

Let us consider the following aspect of the
confinement problem in QCD regarding the
fact that quarks and gluons are not observable single
particles. One heuristic argument why
we can not observe the color single states is
the following \footnote[1]{author
acknowledges Y.M. Cho for elucidating this argument.}:
quarks and gluons are not gauge
invariant and we have no a conserved color charge
like the electric charge in Maxwell theory.
So that quarks and gluons can not be observable
as single physical particles. 
If we accept the hypothesis that a Lorentz gauge model of
gravity with torsion possesses two types of gauge symmetry (\ref{eqI},\ref{eqII})
then we will be forced to accept the confinement
of torsion.
\section{Minimal model of quantum gravity with torsion}

We are interested in such a Lagrangian in Riemann-Cartan space-time which is reduced to
Gauss-Bonnet topological invariant in the limit of Riemannian geometry.
So that, we will consider the following Lagrangian
\bea
{\cal L}&=&-\dfrac{1}{4}(\alpha R_{abcd}^2+(1-\alpha) R_{abcd}R^{cdab} -4 \beta R_{bd}^2 \nn \\
&&     -4(1-\beta) R_{bd}R^{db} +R^2+6 \gamma A_{abcd}^2), \label{startL}
\eea
where the irreducible tensor $A_{abcd}$ is defined as follows \cite{hayashiI}
\bea
A_{abcd}=\dfrac{1}{6} (R_{abcd}+R_{acdb}+R_{adbc}+R_{bcad}+R_{bdca}+R_{cdab}).
\eea
It turns out that the model described by the Lagrangian (\ref{startL})
admits dynamical degrees of torsion (contortion) for the special values of the parameters
$\beta=0, \gamma=-3\alpha$.  The parameter $\alpha$ provides
unimportant overall factor, so that one can set $\alpha=1$ without loss of generality.
For convenience, we will keep the parameters $\alpha, \gamma$ arbitrary and later
we will show that propagating torsion requires a unique value for the parameter $\gamma=-3$.
The Lagrangian (\ref{startL}) takes the form (omitting total divergence terms)
\bea
{\cal L}=\difrac{1}{2}\Big[R_{abcd}^2+2R_{abcd}R^{cdab}
+6 R_{abcd}R^{acdb} \Big] \label{lastL}
 \eea

In addition to the equations of motion 
$\delta {\cal L}/\delta K_{bcd}=0$
one should
impose gauge fixing conditions. To fix Lorentz gauge
symmetry we choose the following constraints  which
are compatible with equations of motion
\bea
&& \pro^i (K_{i0\delta}-K_{\delta 0i})=0, \label{G1}\\
&&(\alpha+\gamma) \pro^i K_{i\gamma\delta}=\gamma (\pro^i K_{\gamma i\delta}
-\pro^i K_{\delta i\gamma}),        \label{G2} \\
&& \pro^i \pro^j K_{i0j}=0. \label{G3}
\eea
For simplicity we choose the covariant constant background space-time
as a Riemannian space-time of constant curvature
$\hat R_{abcd}= \rho (\eta_{ac} \eta_{bd}-\eta_{ad}\eta_{bc})$.
We will use the following decomposition of the space components of the contortion
\bea
&&K_{\mu\gamma\delta}=\epsilon_{\gamma \delta \rho}\stackrel{\ast}{K}_{\mu\rho}, \nn \\
&&\astK_{\mu\rho}=\TTS_{\mu\rho}+\difrac{1}{2}(\delta_{\mu\rho} \Delta-\pro_\mu\pro_\rho)\TS
 +(\pro_\mu S_\rho+\pro_\rho S_\mu) +\epsilon_{\mu\rho\sigma} A_\sigma , \nn \\
 &&K_{\mu 0 \rho}=\TTR_{\mu\rho}+\difrac{1}{2}(\delta_{\mu\rho} \Delta-\pro_\mu\pro_\rho)\TR
 +(\pro_\mu R_\rho+\pro_\rho R_\mu) +\epsilon_{\mu\rho\sigma} Q_\sigma,
\eea

Some of the
equations of motion and gauge conditions
represent constraints for components of $K_{bcd}$.
One can solve all constraints and gauge conditions in linearized approximation
and substitute solution into the initial Lagrangian.
After lengthy calculations one can find a final
effective Lagrangian (quadratic in fields $K_{bcd}$)
which contains only independent physical dynamical degrees of freedom
\bea
\difrac{1}{2} {\cal L}_{eff}^{(2)}&=&
\difrac{3}{8}\TTS^{\alpha\beta} \difrac{\rho}{\Delta} \Box \TTS_{\alpha\beta}
+ A^{tr \alpha} (\Box +2 \rho) A^{tr}_\alpha \nn \\
&&-(\varphi +\pro_0 \psi)^2 - \varphi \difrac{\Box+6 \rho}{\Delta} \varphi
+\psi (\Box +6 \rho) \psi, \\
&&\varphi=\pro^i Q_i, ~~~~~~~~~~~ \psi=-\difrac{2}{3}\pro^i S_i.
\eea

In conclusion, we propose a
model of quantum gravity with dynamical torsion.
The model has a number of advantages to compare
with Yang-Mills type Lorentz gauge gravity.
In the absence of torsion the model reduces to a pure
topological gravity, i.e., one has a topological phase
where the metric is not specified a priori. The metric
can obtain dynamical content after
dynamical symmetry breaking in the phase of
effective Einstein gravity which is induced by quantum
torsion corrections.
Remarkably, the contortion in our model has the same number of
degrees of freedom as the metric in Einstein gravity.
This could be an additional hint that torsion can be interpreted as a quantum counter-part to
the classical graviton. 

Author acknowledges Organizing  Committee of the Symposium SSP 2009
for the invitation and hospitality. Author thanks Professor Y. M. Cho
and Professor K.-I. Kondo for numerous interesting
discussions.


\begin{thebibliography}{99}
\bibitem{uti} R. Utiyama, Phys. Rev. {\bf 101}, 1597 (1956).
\bibitem{kibble} T.W.B. Kibble, J. Math. Phys. {\bf 2}, 212 (1961).
\bibitem{cho11} Y.M. Cho, Phys. Rev {\bf D14}, 2521 (1976).
\bibitem{ivan1} D. Ivanenko and G. Sardanashvily,
Phys. Rep. {\bf 94}, 1 (1983).
\bibitem {hehl}  F.W. Hehl, J.D. McCrea, E.W.  Mielke, and Y. Ne'eman,
Phys. Rep. {\bf 258}, 1 (1995).
\bibitem{pak} D.G. Pak and S.W. Kim, Class. Quant. Grav. {\bf 25} (2008) 065011.
\bibitem{cho1} Y.M. Cho, Phys. Rev. {\bf D21}, 1080 (1980);
Phys. Rev. {\bf D23}, 2415 (1981); Phys. Rev. Lett. {\bf 46}, 302 (1981).
\bibitem{periw} V. Periwal, hep-th/9808127.
\bibitem{fadd1} L. Faddeev, A. J. Niemi, Phys.Lett. {\bf B449} 
(1999) 214.
\bibitem{fadd2} L. Faddeev and A. Niemi, Nature {\bf 387}, 58
(1997).
\bibitem{hayashiI} K. Hayashi and T. Shirafuji, Progr. of Theor. Phys. {\bf 64} (1980), 866.
\end{thebibliography}

\end{document}